\documentclass[AMA,STIX1COL,natbib]{WileyNJD-v2}

\usepackage{NJDnatbib}
\usepackage{graphicx}
\usepackage{amsmath,amssymb,amsthm}
\usepackage{mathtools}
\usepackage{verbatim}
\usepackage{caption}
\usepackage{subcaption}

\usepackage{moreverb}

\numberwithin{equation}{section}

\newcommand{\E}{\mathbb{E}}

\newcommand{\Z}{\mathbb{Z}}
\newcommand{\N}{\mathbb{N}}

\newcommand{\ind}{\mathbf{1}}

\DeclareMathOperator{\Var}{Var}

\newcommand\BibTeX{{\rmfamily B\kern-.05em \textsc{i\kern-.025em b}\kern-.08em
T\kern-.1667em\lower.7ex\hbox{E}\kern-.125emX}}

\articletype{PREPRINT}%

\received{}
\revised{}
\accepted{}
\copyyear{2025}
\startpage{1}

\begin{document}

\title{AutoWMM and JAGStree - R packages for Population Size Estimation on Relational Tree-Structured Data}

\author[1,2]{Mallory J Flynn}

\author[1]{Paul Gustafson}

\authormark{Mallory J Flynn \textsc{et al}}

\address[1]{\orgdiv{Department of Statistics}, \orgname{University of British Columbia}, \orgaddress{\state{Vancouver, BC}, \country{Canada}}}

\address[2]{\orgname{British Columbia Centre for Disease Control}, \orgaddress{\state{Vancouver, BC}, \country{Canada}}}

\corres{Mallory Flynn, \email{mallory.flynn@stat.ubc.ca}}

\presentaddress{Department of Statistics, Faculty of Science, University of British Columbia, 2207 Main Mall, Vancouver, BC, V6T 1Z4, Canada}

\abstract[Abstract]{The weighted multiplier method (WMM) is an extension of the traditional method of back-calculation method to estimate the size of a target population, which synthesizes available evidence from multiple subgroups of the target population with known counts and estimated proportions by leveraging the tree-structure inherent to the data.  Hierarchical Bayesian models offer an alternative to modeling population size estimation on such a structure, but require non-trivial theoretical and practical knowledge to implement. While the theory underlying the WMM methodology may be more accessible to researchers in diverse fields, a barrier still exists in execution of this method, which requires significant computation.  We develop two \texttt{R} packages to help facilitate population size estimation on trees using both the WMM and hierarchical Bayesian modeling;  \textit{AutoWMM} simplifies WMM estimation for any general tree topology, and \textit{JAGStree} automates the creation of suitable JAGS MCMC modeling code for these same networks.}


\jnlcitation{\cname{%
\author{M.J. Flynn}, and
\author{P. Gustafson}} (\cyear{2025}), 
\ctitle{AutoWMM and JAGStree - R packages for Population Estimation on Tree-Structured Data}, \cjournal{Preprint}, \cvol{2025;02:1--18}.}

\maketitle

\footnotetext{\textbf{Abbreviations:} MM, multiplier method; WMM, weighted multiplier method}

\section{Introduction}\label{intro}

Population size estimation is of interest in many disciplines.  Key populations may be partially hidden from data for a variety of reasons, yet the magnitude of these populations is important to inform policy makers and other stakeholders.  While many factors contributing to undercounting of a given target population are known, the effect of these factors are often complex and multi-faceted.

The \textit{multiplier method} is a commonly applicable back-calculation estimation method, due to it's simplicity and wide applicability using existing data.  In its most general form, a service or unique object identifier is distributed to individuals of a target population, which need not be random.  For example, in public health \cite{abdulqMM}, some service or trait may serve as a unique identifier defining a subpopulation.  The count of individuals receiving this service or unique object is then used with an independent, representative estimate of the proportion of the population receiving the service or unique object, to obtain a size estimate of the target population \cite{fearonMM}.  While the multiplier method makes it possible to exploit existing information to formulate population size estimates, it is sensitive to the accuracy of both marginal counts and proportion estimates, and will not produce reliable or robust population size estimates with poor quality or variable data \cite{hickmanCRC2005}. Traditionally, back-calculated estimates generated using the multiplier method have used the count of a single sub-population of the target population; in practice, multiple sub-populations with corresponding estimates of proportions may be available, and the weighted multiplier method (WMM) is a more suitable approach \cite{flynnmethods}.

A fully Bayesian approach also provides a framework capable of generating target population size estimates, and is also used to estimate the impact of intervention strategies or therapies \cite{irvinevarbayes2019,irvinethnkits2018,irvineHIV2018}.  A hierarchical Bayesian model is inherently well suited to synthesizing expert knowledge and data from multiple sources to make inference on model values and parameters \cite{bayespop, tancrediBayes, gustafMM, kingBayes, winBUGS, fienbergBayes}.  While Bayesian methods are more robust to variations in available data or unrepresentative sources, models must often be tailored to the specific problems at hand, requiring non-trivial theoretical and practical knowledge to implement.  Where this is difficult or infeasible, the WMM is an attractive alternative.  The method is simpler to understand and implement than Bayesian modeling, but incorporates several favourable Bayesian attributes.  Since all observed leaf nodes can be paired with the respective branching proportions to generate a root population size estimate, multiple sources of evidence which draw on prior knowledge are similarly synthesized, and the underlying network structure is also leveraged \cite{flynnmethods}.

Here, we develop an \texttt{R} package, called \textit{AutoWMM}, which facilitates ease of use of this methodology, making it simple to implement on any general tree structure.  For estimation for which a hierarchical Bayesian model is required or preferable, one barrier to implementation is the development of modeling code which represents the conditional structure of the tree and the distributions associated with both latent and observed nodes and branches, while satisfying the formatting needs of the underlying MCMC sampling program.  A growing number of resources are being developed to assist with code development \cite{bugsnet}.  Since this topology admits a commonly applicable framework for population size estimation problems, we similarly construct an \texttt{R} package which automates the creation of suitable JAGS MCMC modeling code, called \textit{JAGStree}, removing this barrier to implementation of hierarchical Bayesian models.  These packages may be readily downloaded \cite{AutoWMM, JAGStree} and used for population estimation on tree-structured data. In a companion paper, we describe the underlying theory behind both the multiplier-based methodology and the hierarchical Bayesian model \cite{flynnmethods}. In another companion paper, we use the methodology to estimate the number of opioid overdoses events in the province of British Columbia, Canada, over a specified calendar period \cite{flynnapplication}.

\section{Root Population Size Estimation on Trees} 

Suppose multiple subsets of a target population have known or estimated size and are mutually exclusive.  Then a tree can be constructed using the known data, with the root representing the target population, leaves as sub-populations of the root, and possible inner path nodes further describing nested subgroups.  Nodes may be partially or wholly observed, with directed edges associated with the proportion of the parent node population belonging to the child node population.  The tree structure can then be exploited to combine all observed evidence to generate an optimal estimate of the target population, given the constraints of the data.  

\subsection{Overview of Bayesian Methodology and JAGS}
Suppose we are interested in the distribution of a parameter or hypothesis, $\theta$, given a set of observed data, $\mathcal{D}$.  By incorporating past knowledge about $\theta$ through a choice of \textit{prior distribution}, $p(\theta)$, conditional on some set of hyperparameters $\alpha$, an application of Bayes theorem dictates that
\[
p(\theta|\mathcal{D}) = \frac{p(\mathcal{D}|\theta)\cdot p(\theta)}{p(\mathcal{D})},
\] 
where $p(\mathcal{D}|\theta)$ is the likelihood of the observed data and $p(\theta|\mathcal{D})$ is the \textit{posterior distribution} \cite{robertbayesbook}.  The \textit{marginal likelihood}, $p(\mathcal{D})$, is often difficult to compute as the integral cannot necessarily be solved in closed form in the continuous case, while in the discrete case, it may involve summing over infinitely many values $\theta$.  Fortunately, it does not depend on $\theta$; thus a commonly used representation of the posterior distribution is through the expression 
\begin{equation}\label{bayesthm}
p(\theta | \mathcal{D}) \propto p(\mathcal{D} | \theta) \cdot p(\theta).
\end{equation}

A closed form of $p(\theta|\mathcal{D})$ is often not obtainable, and numerical approximation techniques are used in place of exact solutions.  Typically, Bayesian methods are implemented using Markov chain Monte Carlo (MCMC) sampling, where a Monte Carlo random simulation is performed using a Markov chain to explore the state space.    
After specifying the model and a set of initial conditions by processing observations and choosing hyperparameters, MCMC can then be used to obtain posterior distributions on parameters from the model for estimation purposes by constructing a Markov chain with the posterior distribution as the steady-state distribution \cite{murphy}.  One choice for implementation is JAGS \cite{jags}, a Gibbs sampler using MCMC simulation designed to analyze hierarchical Bayesian models. JAGS employs the Metropolis algorithm to create a Markov chain with the desired distribution, $p(X)$, as its stationary distribution.  The initialized density is corrected through acceptance or rejection of proposal steps and ultimately generating a series of random values representative of the posterior under the assumption of convergence.  Gibbs sampling, in particular, cycles through each parameter of interest, determining conditional posteriors of each by using current values of the others, and accepts proposed values from the posterior with probability one \cite{robertbayesbook}.  This provides a series of low dimensional steps in simulation which still converge to the correct stationary distribution, $p(X)$ \cite{robertbayesbook}.  Where this sampling strategy is not possible or unfeasible, a number of other sampling techniques exist (which JAGS automatically resorts to, when needed \cite{jags}).  For instance, Gibbs sampling is a special case of Metropolis-Hastings, a more general algorithm that employs a proposal distribution, $q$, as part of the acceptance-rejection scheme.  Once a new sample, $x'$, has been proposed, the decision to accept or reject this proposal is made according to a formula, whose construction ensures that the proportion of time spent in each state is proportional to the steady state distribution, $p$ \cite{murphy}.  By calculating
\[
\alpha = \frac{p(x')\cdot q(x|x')}{p(x)\cdot q(x'|x)},
\]
we accept the proposal with probability $r$ \cite{murphy}, where
\[
r = \min(1, \alpha).
\]
It can be seen from the formula for $\alpha$, that $p$ need only be proportional to the target distribution, as normalizing constants in the numerator and denominator cancel.

\begin{figure}
	\centering
	\includegraphics[height=0.5\textheight]{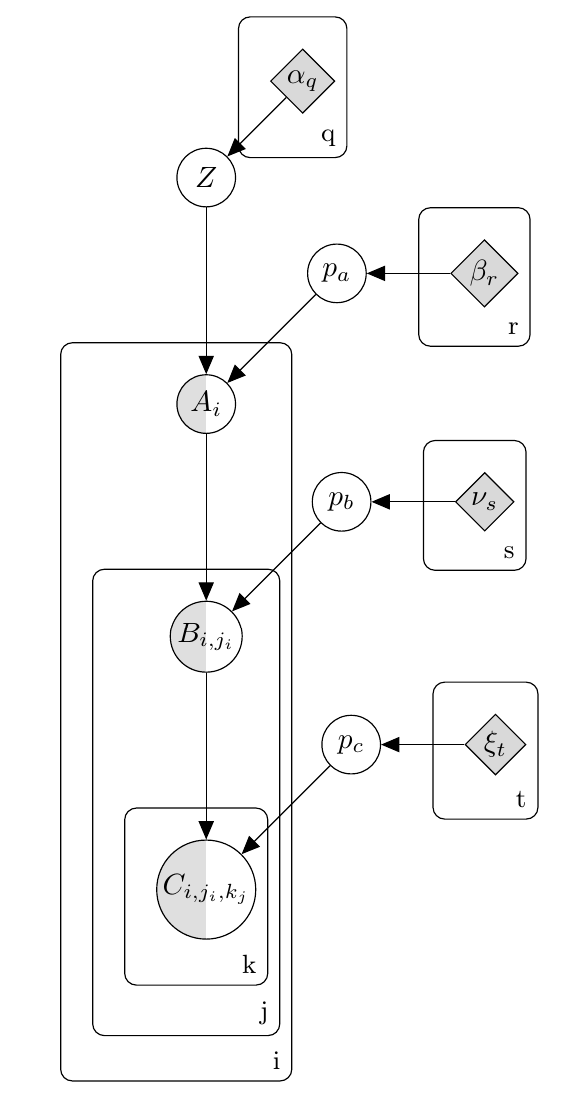}
	\caption{Graphical model for general tree-structured data with three levels. Shaded nodes represent parameters derived from prior knowledge or literature, with half-shaded nodes representing that data may or may not be available at these nodes, depending on the indices.  Indices $i,j_i,k_j \in \N$, where $j,k$ differ depending on the number of child nodes of each parent.  The dimensions of hyperparameters $q,r,s,t \in \N$ differs also based on the chosen distributions, but we assume these are known hyperparameter inputs to the priors of $Z$, $p_a$, $p_b$, and $p_c$.}
	\label{fig:generalDAG}
\end{figure}

Joint distributions of random variables are often represented using a directed, acyclic graph (DAG) as a graphical model representing conditional dependence assumptions.  An ordered Markov property exists between nodes and their parent, such that a random variable represented by any node depends only on the random variable represented by it's immediate parent, and not any higher predecessors \cite{murphy}.  A general DAG for a tree with three levels can be found in Figure \ref{fig:generalDAG}. 

\subsection{Overview of WMM}\label{sec:overviewWMM}

The WMM is described at length in Flynn \& Gustafson \cite{flynnmethods}.  In brief, the traditional multiplier method serves as the building blocks behind a method of synthesizing evidence across a number of subgroups, each capable of generating a target population size estimate.  By constructing a tree, with a root node representing the target population and leaf nodes representing subgroups with known marginal counts, an estimator is formed by a weighted average utilizing variance-minimizing weights. 
More formally, let $\mathcal{T} \equiv \mathcal{T}_Z(V, E)$ be the tree-structured data on which we wish to make inference.  Let $V(\mathcal{T})$ and $E(\mathcal{T})$ be the sets of nodes, $v$, and edges, $e$, in $\mathcal{T}$, respectively, where an edge $e$ is an ordered pair of nodes.  We define a path between two nodes $v_0$ to $v_K$, to be a sequence of edges $\gamma(v_0, v_K) \subseteq E(\mathcal{T})$ connecting $v_0$ to $v_K$.
We further define a substructure, $\mathcal{T}^D \equiv \mathcal{T}_Z^D(V^D, E^D) \subseteq \mathcal{T}$, which represents those nodes and edges of $\mathcal{T}$ for which we have data, so that all nodes and edges in $\mathcal{T} \setminus \mathcal{T}^D$ are latent.  
When population flows through a tree-like data structure and multiple counts are available for a subset of the leaves, then so long as branching probability estimates are available for each root-to-leaf path, this scenario admits multiple back-calculated estimates of the root node.  Such paths can be formalized by the following definition:
\begin{definition}\label{def:informativepath}
	(Informative path) Let $\mathcal{T}_Z(V,E)$ be a tree with nodes $V$ representing populations, root $Z \in V$, and edge set $E$.  Let $\mathcal{L} \subset V$ be the set of leaves of $\mathcal{T}_Z$.  A path $\gamma(Z, L) \subseteq E$, is called \textit{informative} if all the following conditions are true:
	\begin{itemize}
		\item[(i)] $L \in \mathcal{L}$,
		\item[(ii)] an estimate $D_L$ exists for the marginal count of $L$, and
		\item[(iii)] there exists an estimate for the branch probability $p_e$, $\forall e \in \gamma(Z,L)$.
	\end{itemize}
	We further define the set $\mathcal{L^*}$ to be 
	\[
	\mathcal{L^*} = \{L \in \mathcal{L}: \gamma(Z,L) \text{ is \textit{informative}}\}.
	\]
\end{definition} 
We let $\mathcal{C}^D(\mathcal{T})$ denote the set of all combinations of branch estimates available to inform $E^D(\mathcal{T})$.
For each $C \in \mathcal{C}^D(\mathcal{T})$, we sample $M$ sets of branching probabilities from distributions determined by the prior knowledge of edges $e \in E^D(\mathcal{T})$.  Each set of samples, indexed by $m$, is combined with marginal leaf counts of each $L \in \mathcal{L^*}$, generating $M$ back-calculated values of the root node per leaf $L$ (not necessarily unique).  This process generates an $M$ by $|\mathcal{L^*}|$ matrix, $\mathbf{M}$, of estimates of the root population size.  In particular, each column of $\mathbf{M}$ represents a leaf, $L$, with observed count, and each row, $m$, corresponds to one sampled realization of the tree, so that a matrix value $M_{m',L'}$ is the back-calculated estimate of the root population given the count at $L'$ and the subset of relevant path probabilities sampled on run $m'$ which are required to perform the back-calculation.  

Population values in $\mathbf{M}$ are log-transformed before calculating the covariance matrix, producing the matrix $\mathbf{L}=\log\mathbf{M}$, the element-wise transformation of $\mathbf{M}$.  This stabilizes the calculation of the weights, $\mathbf{w} \in \mathbb{R}^{|\mathcal{L^*}|}$, where we impose the constraint $\sum_{L \in \mathcal{L^*}} w_L = 1$.  We can then calculate a root population size estimate, 
\begin{equation}\label{eq:wmmL}
\hat{\theta} = f(\mathbf{L} \cdot \mathbf{w}),
\end{equation}
where $(\mathbf{L} \cdot \mathbf{w}) \in \mathbb{R}^{M \times 1}$ and $f$ represents the uniform average.  To obtain the final estimate, we set $\hat{Z} = \exp(\hat{\theta})$. 

Under log-transformation, weights $\mathbf{w}$ are generated using $\mathbf{L}$ and $\hat{Z}$ is a multiplicative function of path estimates raised to the power of the weights $w_i$.  In particular, we have
\begin{align*} 
	\hat{\theta} &=  \sum_m  \left[ \sum_L \frac{\mathbf{w}_L}{M} \cdot \log( \mathbf{M}_{m,L})\right]\\
	&= \log\left(\prod_m \prod_L \mathbf{M}_{m,L}^{\mathbf{w}_L/M} \right)
\end{align*}
and
\begin{equation}\label{eq:Zhat}
\hat{Z} = \prod_{m=1}^M \prod_{L \in \mathcal{L^*}}  \mathbf{M}_{m,L}^{\mathbf{w}_L/M}.
\end{equation}

The above procedure holds for a single set of estimates $C \in \mathcal{C}^D(\mathcal{T})$, which suggests we have only one set of values informing our branching estimates. When $|\mathcal{C}^D(\mathcal{T})|>1$, at least one branch has more than one plausible estimate.  Where more than one previous data source is available to inform any branching estimates, and it may not be immediately clear which estimate should be used. For $|\mathcal{C}^D(\mathcal{T})|>1$, a two-stage weight generating process could instead be used.  For each $C$, we proceed as above up to the point of generating weights $\mathbf{w}_C$, now dependent on the set $C$.  We then generate a vector of estimates, $\hat{\theta}_C \in \mathbb{R}^M$, defined by 
\[
\hat{\theta}_C = \mathbf{L}_C\cdot \mathbf{w}_C.
\]
The process is repeated for each $C \in \mathcal{C}^D(\mathcal{T})$, and estimates are combined to form the matrix $\hat{\Theta} \in \mathbb{R}^{M \times |\mathcal{C}^D(\mathcal{T})|}$.  The covariance matrix of $\hat{\Theta}$ can then be used to generate weights $\mathbf{W}$ which account for total variance among the possible sets, $C$, of branch data.  A final scalar estimate, $\hat{\psi}$, is then given by
\[
\hat{\psi }= f(\hat{\Theta} \cdot \mathbf{W}),
\]
and can be converted to a population estimate at the original scale by setting $\hat{Z} = \exp(\hat{\psi})$. 

The inclusion of prior knowledge in the form of branching distributions imparts a Bayesian element to the WMM methodology, as does the assignment of a distribution to the root population size estimator, being based on a sum of multiplicative terms involving these random variables.  However, some important differences do exist between the two methodologies which do not qualify it as a fully Bayesian approach, and thus suitability of the WMM methodology depends on the inherent assumptions and the available data of the application at hand \cite{flynnmethods}. 

\subsection{On the Subject of Error}
The weighting of informative paths is an important component to synthesizing the data to obtain a root population size estimate in the WMM, and is a reflection of relative path uncertainty.  Assigning distributions to branch probabilities (in place of using fixed probabilities as in traditional back-calculation) accounts for some sources of error, though the errors which are accounted for depend on the methods by which these distributions are constructed.  For example, by directly utilizing survey values as parameters in $Dirichlet$ branching distributions, the sampling procedure will generate variability in root population size estimates as a function of survey sample size, and path-specific estimates can then be compared by their variability as it relates to sample size limitations.  Alternatively, by setting branch distributions subjectively, one may account for errors due to other reasons, such as non-representative samples, though a quantitative method of doing so is not developed herein.  Other errors, such as those due to leaf counts, would be more naturally accounted for through a methodological extension which assigns distributions to node counts as well.  
Bayesian methods can also deal with uncertainty in node counts through the assignment of priors to these counts or the inclusion of ``data uncertainty'' nodes representing missed counts \cite{flynnapplication}.  

\section{Software for Root Population Size Estimation on Trees}

\begin{figure}
	\centering
	\includegraphics[width=0.4\linewidth]{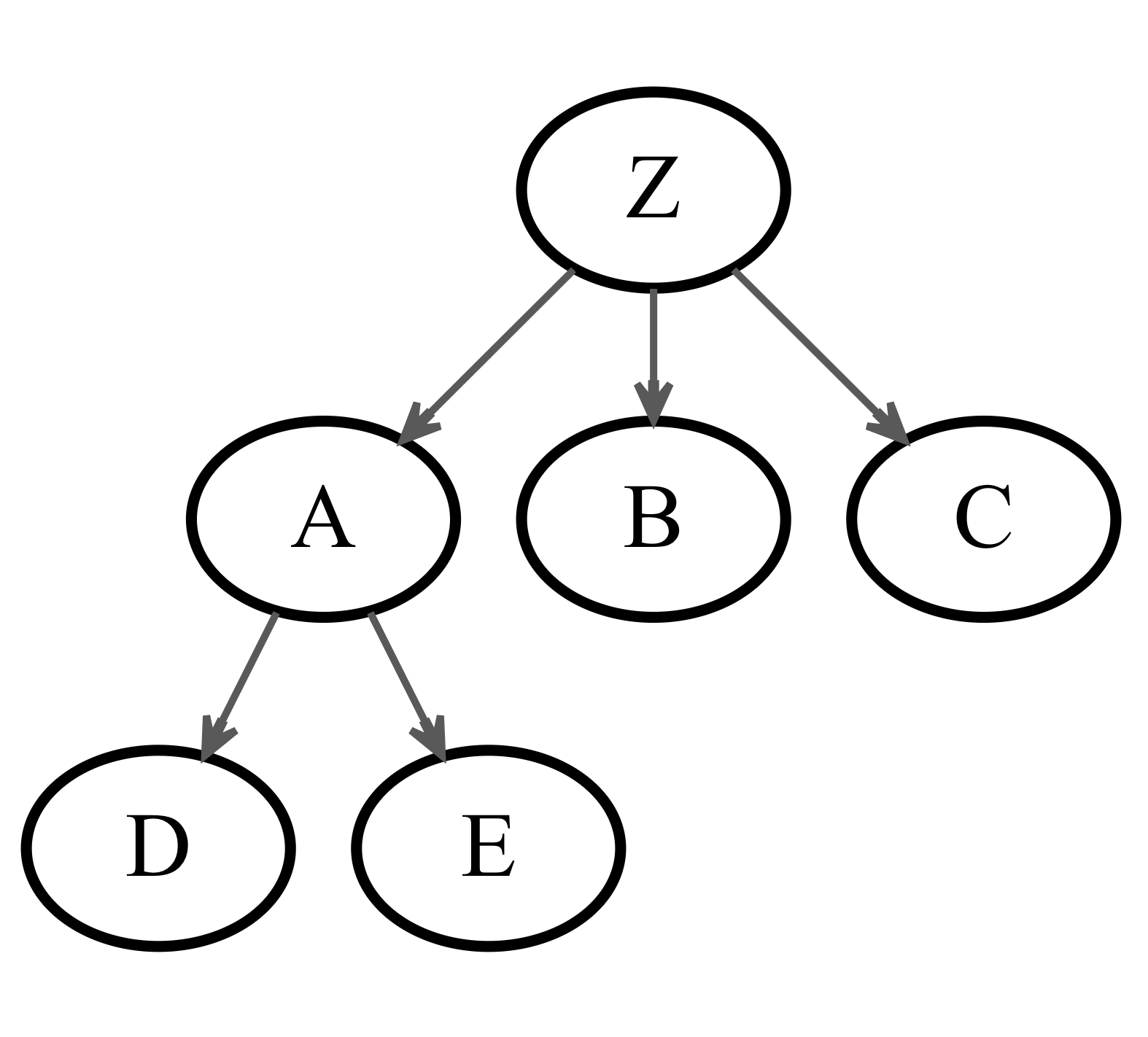}
	\caption{A simple tree use to illustrate the functions and output of the \texttt{R} packages $AutoWMM$ and $JAGStree$, in Sections \ref{sec:autowmm} and \ref{sec:autoJAGS}.}
	\label{fig:autotree}
\end{figure}

Synthesizing multiple sources of information with the WMM or Bayesian modeling enables better integration of all known information into a population estimate.  Unfortunately, more complex techniques are often not as accessible as simpler methods, such as traditional single-source back-calculation, despite their broad range of applications and adaptability to problems in a variety of settings.  
When data are linked as a tree-like structure, some conditions and assumptions placed on this structure allow for automated model generation and analysis.  In the following, we summarize these assumptions, which are likely to be satisfied across a variety of practical applications.  We present two \texttt{R} packages; \textit{AutoWMM}, which may facilitate more widespread use of the WMM, and \textit{JAGStree}, a means to generate Bayesian modeling code for appropriate tree-structured data, minimizing the barriers to implementing such an approach.  The implementation is developed generally, and then illustrated as applied to a simple tree, as in Figure \ref{fig:autotree}.

\subsection{Automating WMM Implementation in \texttt{R}}\label{sec:autowmm}
While traditional back-calculation is relatively simple to execute, the implementation of the WMM requires that a source-dependent sampling procedure be generated to synthesize evidence across the tree's branching structure \cite{flynnmethods}.  Path-specific samples must then be used to calculate weights, a final root estimate, and appropriate confidence intervals.  This section outlines these procedures and encapsulates them into a package for \texttt{R} which automates the WMM estimation process on general tree-structured data.

\subsubsection{\textit{AutoWMM} Overview and Model Assumptions}\label{sec:autowmmtheory}
We assume a basic structure to the data, which is that the flow of information between sources can be represented by a well-defined tree.  We further assume that time is not a considerable factor and does not cause a reporting lag (see section \ref{sec:discussion}).  To implement the WMM, we must have at least one marginal leaf count, and root-to-leaf paths ending in leaves with available counts are informative, as in Definition \ref{def:informativepath}.  

Within any application, a number of possible sources may inform any given set of sibling branches, resulting in several branch sampling approaches in the implementation of the WMM. In the case that all branches of an informed path $\gamma(Z,L)$ are sampled from $Beta$ distributions on binary branching groups, a closed form solution for the expected population size estimate, $\E(\theta_L)$, of root $Z$ exists in some circumstances; for margin count $D_L$ at leaf $L$, we have
\begin{align*}
	\E(\theta_L) 
	&= D_L \cdot \prod_{e \in \gamma(Z,L)} \E(p_e^{-1}),
\end{align*}
where $p_e \sim Beta(\alpha_e,\beta_e)$ and we assume independence of branching distributions.  By the definition of the $n$-th moment of a random variable $X$ \cite{durrett}, when $X\sim Beta(a,b)$ and $a+n >0$, we have
\begin{align*}
	\E(X^{n}) &= \frac{1}{B(a,b)} \int_0^1 x^{n} x^{a-1} (1-x)^{b-1} dx
	= \frac{\Gamma(a+b) \Gamma(a+n)}{\Gamma(a)\Gamma(a + b+ n)},
\end{align*}
and substituting $n=-1$ and the parameters of $p_e$, we have
\begin{equation}\label{eq:betacf}
	\E(\theta_L) = D_L \cdot \prod_{e \in \gamma(Z,L)} \frac{\Gamma(\alpha_e+\beta_e) \Gamma(\alpha_e-1)}{\Gamma(\alpha_e)\Gamma(\alpha_e + \beta_e -1)}.
\end{equation}
Similarly, when each $e \in \gamma(Z,L)$ is sampled from a $Dirichlet$ distribution, a closed-form equation is also available for $\E(\theta_L)$, so long as each $Dirichlet$ distribution describes the entire set of sibling branches.  In this case, for a sibling group $S$, each branch $i \in S$ has associated marginal probability distribution $p_i \sim Beta(\alpha_i, \alpha_T - \alpha_i)$, where $\alpha_T = \sum_{i \in S} \alpha_i$, and the result follows similarly.  In each instance, a closed-form representation of the variance of $\theta_L$ is also obtainable using the moments definition, so that 
\begin{align}\label{eq:pathvar}
	\Var \theta_L 
	&=  \prod_{e \in \gamma(Z,L)} \frac{D_L^2 \Gamma(\alpha_e + \beta_e)\Gamma(\alpha_e -2) }{\Gamma(\alpha_e)\Gamma(\alpha_e + \beta_e -2)} - \prod_{e \in \gamma(Z,L)} \frac{D_L \Gamma(\alpha_e + \beta_e)\Gamma(\alpha_e -1)}{\Gamma(\alpha_e)  \Gamma(\alpha_e + \beta_e -1)}.
\end{align}

While exact solutions bypass sampling and increasing computational efficiency, in practice, sibling branch knowledge is likely to be derived from mixed sources.  This results in an implementation in which sampled values are subjected to rejection schemes or re-weighted using importance sampling.  For instance, if at least one branch of a sibling group is not part of an informative path, then $Dirichlet$ sampling will be used in the case that a single survey source informs all branches of the group which are part of informative paths; if all sibling branches which are part of informative paths have different surveys informing them, then independent $Beta$ distributions will be used, which are constrained by an accept-or-reject algorithm that guarantees a viable joint sample.  
The latter specification aligns with the resulting posterior distribution if we approach branch sampling as a Bayesian process.  

Various cases can be categorized as provided full knowledge for a sibling group rather than a subset of siblings, and either using a single source for such knowledge, or multiple sources. Cases within each of these latter two scenario can be approached as follows.

\noindent\textbf{1. Subset of Informed Sibling Branches}

\begin{figure}
	\centering
	\includegraphics[width=.5\linewidth]{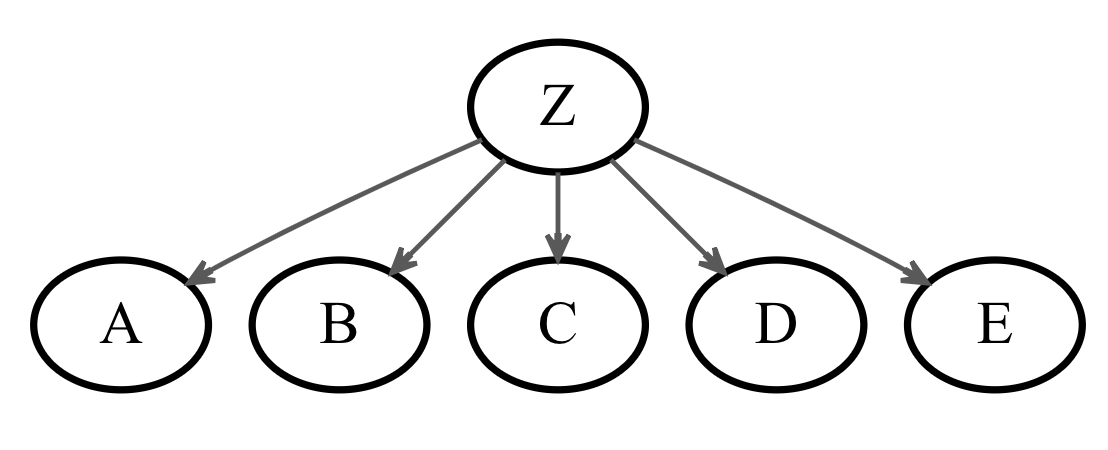}
	\caption{Basic WMM example tree.}
	\label{fig:autowmmfig}
\end{figure}

We first consider when only a subset of branching to members of any particular sibling group is informed. Consider Figure \ref{fig:autowmmfig}, which depicts the simple scenario in which a parent node, $Z$, has five children, $i \in \{A,B,C,D,E\}$, with $p_i$ denoting the respective proportions of the population at $Z$ moving to child $i$.  
Suppose marginal counts exist for only $A$, $B$, $C$.  If a survey exists which examines a representative sample of a population identical to that at node $Z$ and the proportions of this sample which move to the states defined by nodes $i$, then this survey provides all the information required by the WMM to inform the parameters of a $Dirichlet$ distribution on $(p_A, p_B, p_C, p_D, p_E)$.  Furthermore, for informative paths which are specified in this way at each segment, the resulting distribution may be computed analytically, with mean and variance given as in equations \eqref{eq:betacf} and \eqref{eq:pathvar}.

In many circumstances, however, such clean representations will not exist, and we may have a mix of surveys informing different branches of a sibling group.  For example, let $Z_K$ represent survey populations identical to the population at node $Z$, where $K \in \{A,B,C, D, E\}$.  Suppose one survey shows $a$ individuals move to $A$ from a sample of $Z_A$ of size $N_a$, so that $N_a-a$ must move to $B \cup C \cup D \cup E$.  In addition, consider a second survey which shows that $b$ individuals move to $B$ from a sample of size $N_b$ of $Z_B$, so that $N_b - b$ must move to $A \cup C \cup D \cup E$.  Lastly, suppose a third survey shows that $c$ individuals from a sample of size $N_c$ of $Z_C$ move to $C$, so that $N_c - c$ move to one of $A \cup B \cup D \cup E$.  We assume all surveys are mutually independent, and $p_D$, $p_E$ are uninformed (and not required, as marginal counts of $D$ and $E$ are unavailable).  If we construct a uniform prior on the branching probabilities, $(p_A, p_B, p_C, p_D, p_E) \sim Dir(1,1,1,1,1)$, then by multiplying by the likelihoods of the survey data for each survey and a $Beta(1,1)$ distribution for $p_D$ (which is specified only by the prior), we have a posterior density in the $(p_A, p_B, p_C, p_D)$ parametrization, with respect to the Lebesgue measure on a subset of $(0,1)^4$, which is proportional to 
\begin{align*}
&p_A^{(a+1)-1}(1-p_A)^{(N_a-a+1)-1}\cdot p_B^{(b+1)-1}(1-p_B)^{(N_b-b+1)-1} 
\cdot p_C^{(c+1)-1}(1-p_C)^{(N_c-c+1)-1} \cdot p_D^{1-1}(1-p_D)^{1-1}\ind_{\{p_A+p_B+p_C + p_D<1\}}, 
\end{align*}
and simplifies to
\[
p_A^{a}(1-p_A)^{N_a-a}\cdot p_B^{b}(1-p_B)^{N_b-b} \cdot p_C^{c}(1-p_C)^{N_c-c} \cdot \ind_{\{p_A+p_B+p_C + p_D<1\}}.
\]
Under these circumstances, the resulting posterior is not $Dirichlet$ but a representation of drawing four independent $Beta$ samples from a truncated distribution with restricted domain.  The computational realization of this truncation is achieved by an all-or-none rejection scheme for $(p_A, p_B, p_C, p_D)$, which restricts points to those on the probability 5-simplex.  

\noindent\textbf{2. Fully Informed Sibling Branches}

The previous example represents only one type of scenario involving mixed sources of sibling branch estimates; other scenarios must be treated differently, and in particular, when the full set of sibling branches is informed, which may happen in a variety of ways.  Consider, first, the case which in which a marginal count is also available for node $E$, and $p_A, p_B, p_C$ are informed by one single survey which shows $a$ individuals move to $A$, $b$ move to $B$, $c$ move to $C$, and $N_1 - a - b- c$ move to $D \cup E$.  Suppose $p_E$ is informed by another survey, which shows $e$ individuals from a sample of size $N_2$ move to node $E$, the remainder of which move to $A \cup B \cup C \cup D$.  Then similar to the previous example, the posterior density in the $(p_A, p_B, p_C, p_D)$ parametrization is proportional to 
\[
p_E^e(1-p_E)^{N_2-e}\cdot p_A^a \cdot p_B^b \cdot p_C^c (1-(p_A+p_B+p_C))^{N_1-a-b-c} \cdot \ind_{\{p_A+p_B+p_C+p_E <1\}}, 
\]
so that a rejection scheme could again realize this scenario.

Another case where the method handles branch sampling differently is when \textit{all} branches are informed.  If only one survey informs all branches, this is straightforward $Dirichlet$ sampling.  But if more than one survey informs the sibling branches, we must take a different approach.  Referring again to Figure \ref{fig:autowmmfig}, suppose all branches are informed by independently conducted surveys which show $k$ of $N_k$ individuals move to node $K$, and $N_k - k$ move to the complement, with $k \in \{a,b,c,d,e\}$, $K \in \{A,B,C,D,E\}$, and $N_k \subseteq Z_K$.  A rejection scheme which constrains sibling branch probabilities to sum to 1 would run indefinitely, since this occurs with probability 0.  Thus we instead take an importance sampling approach, which we demonstrate by example.  Let $f$ be the desired sampling density, which is proportional to 
\[
\prod_{\substack{(K,k) }} p_K^k(1-p_K)^{N_k-k}\cdot \ind_{\{\sum_K p_K =1\}},
\]
where the product is over the pairs $(K,k) \in \{(A,a),(B,b),(C,c),(D,d),(E,e)\} $.
Let $g$ be the proposal distribution.  Without loss of generality, we construct $g$ to be the density in which we sample all branches except $p_E$ and constrain the samples to sum to less than 1.  $p_E$ is then set deterministically to 
\[
p_E = 1 - \sum_{I \in \{A,B,C,D\}} p_I,
\]
so that $g$ is equal to 
\[
\prod_{(K,k) } p_K^k(1-p_K)^{N_k-k}\cdot \ind_{\{\sum_{K \in \{A,B,C,D\}} p_K \leq 1\}}\cdot \ind_{\{p_E = 1-\sum_{K \in \{A,B,C,D\}} p_K \}}.
\]
The distribution, $g$, can be called the \textit{importance distribution}, and in particular, where $g = 0$ we also have that $f=0$.
In practice, this is again realized with a rejection scheme.  Once we obtain samples $p^{(S)} \in [0,1]^5$ from the distribution $g$, we multiply $p^{(S)}$ by the likelihood ratio, $w(p^{(S)})=f(p^{(S)})/g(p^{(S)})$, to weight the samples.  This weighting ensures the Monte Carlo mean is an unbiased estimator of the mean of $f$, and will result in a set of sibling branch probabilities that sums to 1, as desired \cite{impsamp}.  The importance ratio is given by
\begin{align}
w(p^S) &= \frac{f(p^{(S)})}{g(p^{(S)})} \nonumber\\
&= \frac{p_E^e(1-p_E)^{N_e-e}\cdot \ind_{\{ \sum_{K \in \{A,B,C,D,E\}} p_K =1 \}}}{\ind_{\{\sum_{K \in \{A,B,C,D\}} p_K <1\}}\cdot \ind_{\{p_E = 1-\sum_{K \in \{A,B,C,D\}} p_K \}}} \nonumber\\
&= p_E^e(1-p_E)^{N_e-e},
\end{align}
for all accepted samples $p^{(S)}$.  
In real life applications, surveys used to inform branching probabilities may be draws from the actual population $Z$, in place of populations $Z_K$.  If branching surveys are sampled from $Z$, data informing branching probabilities is not independent of node populations, and the above analysis provides an approximation in this case.

\subsubsection{\textit{AutoWMM} Implementation and Examples} \label{sec:autowmmimp}

The automated WMM has been written as the \texttt{R} package ``\textit{AutoWMM}'' using \texttt{R} version 4.1.2 \cite{AutoWMM}.
A specific data structure is required to use the basic functions of the \textit{AutoWMM} package, which include automated root population size estimation using the WMM \cite{flynnmethods} and tree rendering functionality.  In particular, a data frame with at least five columns is required: \textit{from} and \textit{to} columns defining a directed edge between nodes, \textit{Estimate} and \textit{Total} columns derived from prior survey data to establish sampling branch parameters, and a \textit{Count} column which provides marginal leaf counts where known (see Table \ref{table:autowmmdf}). 


\begin{table}[ht]
	\centering
	\begin{tabularx}{\textwidth}{|lcX|}
		\hline \hline
		Column Name & Data Type & Description \\ 
		\hline
		\hline
		\texttt{from} & string & Node label denoting starting node of a directed edge between \texttt{from} node and \texttt{to} node.\\
		\hline
		\texttt{to} & string & Node label denoting endpoint node of directed edge between \texttt{from} node and \texttt{to} node, \\
		\hline
		\texttt{Estimate} & $\Z^+$ & A $Beta$ or $Dirichlet$ parameter for sampling branch probabilities; assumed to come from prior survey, where \texttt{Estimate} number of subjects are observed at node \texttt{to}, from a sample of node \texttt{from} of size \texttt{Total}. \\ \hline
		\texttt{Total} & $\Z^+$ & A $Beta$ or $Dirichlet$ parameter for sampling branch probabilities; assumed to come from prior survey, where \texttt{Estimate} number of subjects are observed at node \texttt{to}, from a sample of node \texttt{from} of size \texttt{Total}. \\ \hline
		\texttt{Count} & $\Z^+$ & Denotes a known marginal leaf count of node \texttt{to}; if node \texttt{to} is not a leaf, or is a leaf but population is unknown, takes the value \texttt{NA}. \\ \hline
		\texttt{Population} (optional) & logical & Optional column in case \texttt{from} and \texttt{to} columns are derived from population-level data; exact probability will be used in place of sampling. \\ \hline
		\texttt{Description} (optional) & string & Optional column describing \texttt{to} node; for use in tree rendering only. \\ \hline
		\hline
	\end{tabularx}
	\caption[Data frame column requirements for \textit{WMM} \texttt{R} package.]{Data frame column requirements for \textit{AutoWMM} \texttt{R} package.}
	\label{table:autowmmdf}
\end{table}

\begin{table}[ht]
	\centering
	\begin{tabularx}{\textwidth}{|lX|}
		\hline \hline
		Function Name: Input $\to$ Output Types & Description \\ 
		\hline
		\hline
		\texttt{makeTree}: data frame $\to$ \texttt{data.tree} structure & Creates a modified \texttt{data.tree} structure from a \texttt{data.frame} object; checks data frame structure to ensure the \texttt{data.tree} object can be used for root node estimation, and generates internal values needed for WMM estimation.\\ 
		\hline
		\texttt{drawTree}: \texttt{makeTree} object $\to$ plot object & A pre-estimation visual depiction of tree.  Node descriptions are given within nodes if provided by \textit{Description} column in dataframe used in \texttt{makeTree}.  Probabilities given by the ratio of \textit{Estimate} over \textit{Total} are given along branches.\\ 
		\hline
		\texttt{weightedTree}: \texttt{makeTree} object $\to$  list, \texttt{makeTree} object & Generates a back-calculated estimates from each leaf with a known marginal count, by sampling branch probabilities and using the multiplier method to generate path-specific estimates, then calculates a variance-weighted average using internal functions.\\ \hline
		\texttt{countTree}: \texttt{makeTree} object $\to$ plot object & For use post-estimation, once \texttt{makeTree} object has been modified with \texttt{weightedTree}.  Generates plot of tree with root estimate displayed in the root, and marginal counts in leaf nodes contributing to the root estimate.  Average of branch samples displayed on each branch. \\ 
		\hline
		\texttt{estTree}: \texttt{makeTree} object $\to$ plot object & For use post-estimation, once \texttt{makeTree} object has been modified with \texttt{weightedTree}.  Generates plot of tree with root estimate displayed in the root node, and the average path-specific root estimates which contributed to the weighted average are displayed in the corresponding leaf node. Average of branch samples displayed on each branch.\\ 
		\hline
		\hline
	\end{tabularx}
	\caption[Summary of functions in the \textit{AutoWMM} \texttt{R} package.]{Summary of functions in the \textit{WMM} \texttt{R} package.}
	\label{table:autowmmfxns}
\end{table}
The package contains several functions for rendering tree diagrams and performing estimation using the WMM.  
The \texttt{makeTree} function is the first point of contact with the rest of the package functionality.  Once a data frame as in Table \ref{table:autowmmdf} has been created describing the data and its tree structure, the \texttt{makeTree} function is used to turn the \texttt{data.frame} object into a \texttt{data.tree} structure, using the imported \textit{data.tree} package \cite{datatree}.  This structure can be further modified through the functions of the \textit{data.tree} package if desired, but some commonly sought plotting features are provided through the \texttt{drawTree}, \texttt{countTree}, and \texttt{estTree} plotting functions in the \textit{AutoWMM} package.  

\begin{figure}
	\centering
	\includegraphics[width=.7\linewidth]{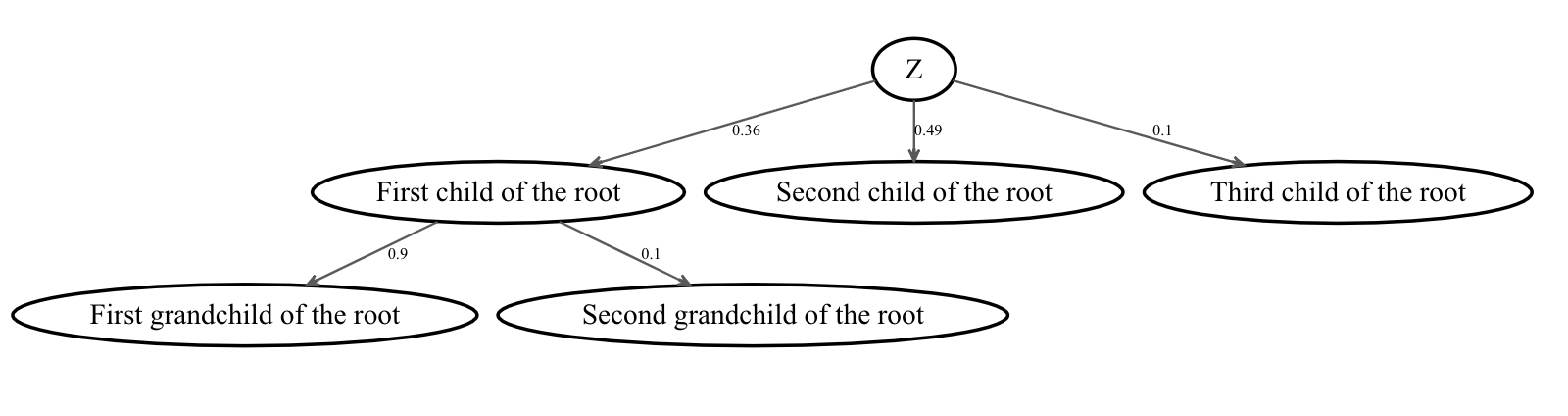}
	\caption{Example of \texttt{drawTree} output of the tree in Figure \ref{fig:autotree}, setting \texttt{probs = TRUE} and \texttt{desc = TRUE}.  \texttt{drawTree} can be used pre-estimation to visualize tree data and prior branching estimates.}
	\label{fig:drawTreeex}
\end{figure}

The \texttt{drawTree} function can be applied to a \texttt{makeTree} object pre-estimation, and generates a plot of the tree structure described by the data.  If a $Description$ column is used, then these summaries can appear in the nodes themselves by the argument \texttt{desc = TRUE}; if not, node labels from the $to$ and $from$ columns will appear within nodes (see Table \ref{table:autowmmdf}).  Probabilities of branches with non-empty $Estimate$ and $Total$ columns can be displayed along edges by the argument \texttt{probs=TRUE}, and reflect the observed proportions given by prior surveys as opposed to sample means.  An example of the tree diagram rendered using this function and based on the example below, can be found in Figure \ref{fig:drawTreeex}.

The \texttt{wmmTree} function is the central function of the package, performing the WMM on the tree with the data provided.  Branching distributions are defined as in Section \ref{sec:autowmmtheory}, and the number of samples drawn is specified through the argument \texttt{sample\_length}.  A variance-minimizing weights matrix is calculated using the inverse covariance matrix of samples $\mathbf{M}$, as in equation \eqref{eq:Zhat}.  Variation among columns can be large if root estimates within each row differ considerably, such as when path distributions depend on non-representative estimates.  Alternatively, when root population size estimates largely agree across informative paths, variation across rows will be small.  The former can result in numerical instability in generating the inverse covariance matrix of $\mathbf{M}$, while the latter can result in near-singularity and problems with matrix inversion.  The \texttt{wmmTree} function handles these issues internally by generating the log transformed matrix $\mathbf{L}$ to provide increased numerical stability, as in equation \eqref{eq:wmmL}, and by using the pseudo-inverse to address near-singularity.
The function formally outputs a list with four entries: the final root population estimate, the rounded root population estimate with corresponding 95\% confidence interval, the vector of log-transformed root estimates of length \texttt{sample\_length}, and the weights vector.  

Under the assumptions on the general tree structure, the true distributions of branching probabilities and node counts are known ($Dirichlet$, $Multinomial$) with unknown parameters.  In this sense, the WMM method can be thought of as a parametric bootstrapping method, with node distributions assumed fixed.  By construction, $(1-\alpha)$\% of the time, the true parameter value for the approximating distribution falls between the $(\alpha/2)$\% and $(1-\alpha/2)$\% quantiles (denoted $\hat{\tau}^*_{\alpha/2}$ and $\hat{\tau}^*_{1-\alpha/2}$, respectively) of the bootstrap replicates $\hat{\tau}^*$ of $\hat{\tau}_n$.  This can be reported as a confidence interval, called the \textit{bootstrap percentile interval} \cite{wasserman}:
\begin{equation}\label{eq:bsconf}
(\hat{\tau}^*_{\alpha/2}, \hat{\tau}^*_{1-\alpha/2}).
\end{equation} 
Other constructions of bootstrap confidence interval may improve on this interval under some conditions.  In particular, the \textit{bootstrap pivot confidence interval}, tries to also take into account the difference between the parameter value in the true distribution and that in the approximation, and is given by \cite{wasserman}:
\[
(2 T_n(\hat{\tau}_n) - \hat{\tau}^*_{1-\alpha/2}, 2 T_n(\hat{\tau}_n) - \hat{\tau}^*_{\alpha/2}),
\]
where $T_n$ is any function of the data.
A precise specification of the confidence interval is unclear for the WMM model.  Given the dependence of the model on the source data, the final root node estimate is produced by a weighted mean of back-calculated estimates given by path-specific bootstrap samples, so that even a bootstrap sample and appropriate confidence interval construction is not immediately clear.  To provide the most straightforward reporting of uncertainty, the default confidence interval produced by the \texttt{wmmTree} function is the central 95\% given by the quantiles, as in equation \eqref{eq:bsconf}:
\[
(\hat{\Theta}^*_{\alpha/2}, \hat{\Theta}^*_{1-\alpha/2}).
\]
An alternate interval, is given by 
\begin{equation}\label{eq:wmmconfint}
(\log(\hat{\Theta}) - 2\sqrt{V}, \log(\hat{\Theta}) + 2\sqrt{V}),
\end{equation}
where $\hat{\Theta}$ is the root population size estimate given by the \texttt{wmmTree} function and $V = 1/\mathbf{e}^T \Sigma \mathbf{e}$, where $\Sigma$ is the precision matrix of the log-estimates of the root population size from each informative path.  The interval in equation \eqref{eq:wmmconfint} can be used to produce uncertainty intervals through the argument \texttt{int.type = var} if desired, though caution should be taken in interpreting results, with comparisons to other intervals offering some insight to how well the interval behaves in the particular context.  The Cox method \cite{lognormconfint} provides a third alternative to generate confidence intervals through the argument \texttt{int.type = cox} in the \texttt{wmmTree} function.  This interval is based on $\Theta = \E(X)$ where $X$ is $LogNormal$:
\[
\left(\overline{\log(\hat{\Theta})} + \frac{s^2}{2} \right) \pm 1.96\sqrt{\frac{s^2}{N} + \frac{s^4}{2(N-1)}}
\]
where $N$ is the \texttt{sample\_length} and $s^2$ is the sample variance of $\log(\Theta)$.    

\begin{figure}
	\centering
	\includegraphics[width=0.4\linewidth]{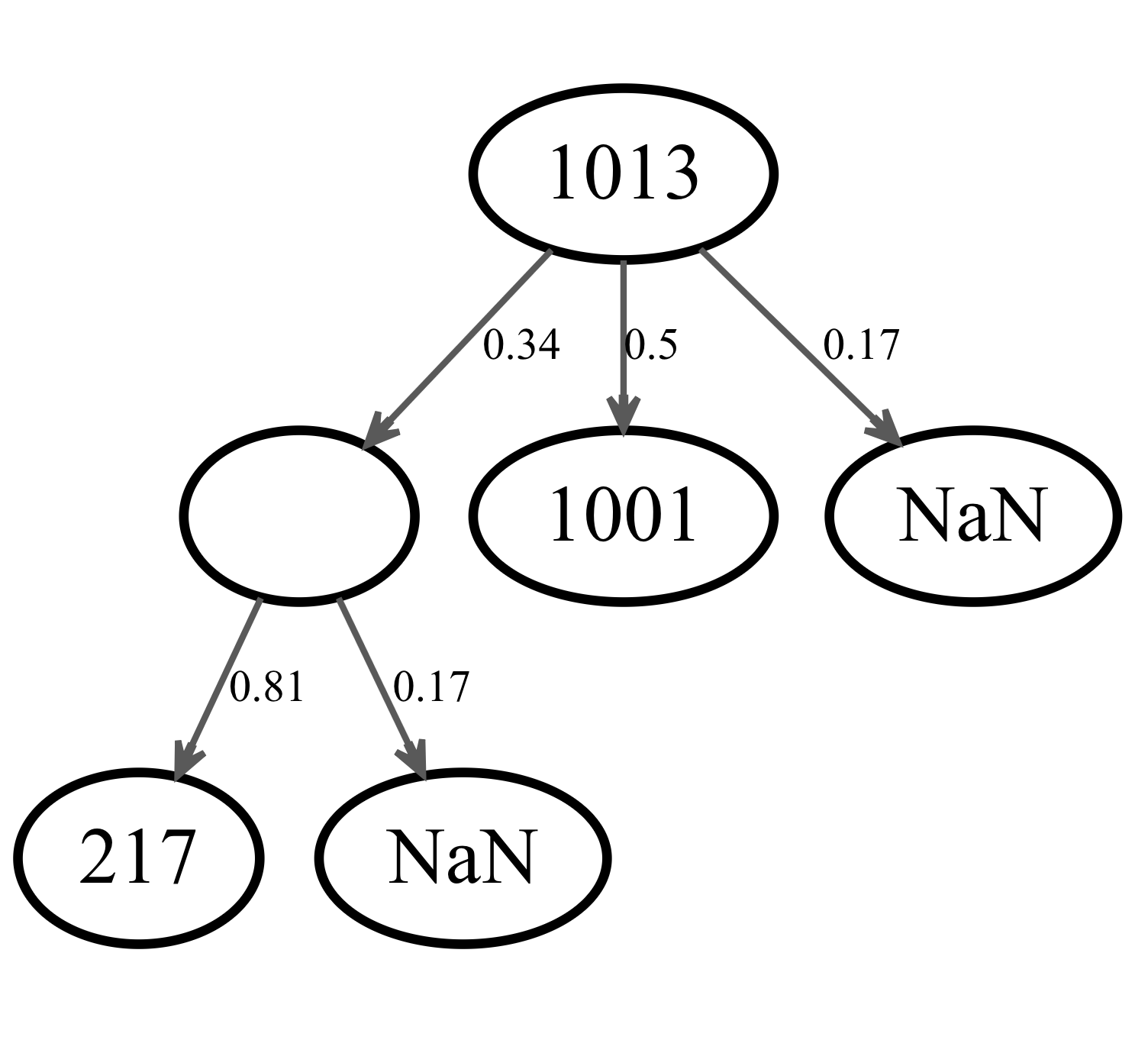}
	\caption{Example \texttt{estTree} output, using the tree defined in Figure \ref{fig:autotree}. \texttt{estTree} is for use post-estimation to visualize tree data with root estimate provided by the WMM, and branch-specific estimates provided in the corresponding leaves.}
	\label{fig:estTreeex}
\end{figure}

\begin{figure}
	\centering
	\includegraphics[width=0.4\linewidth]{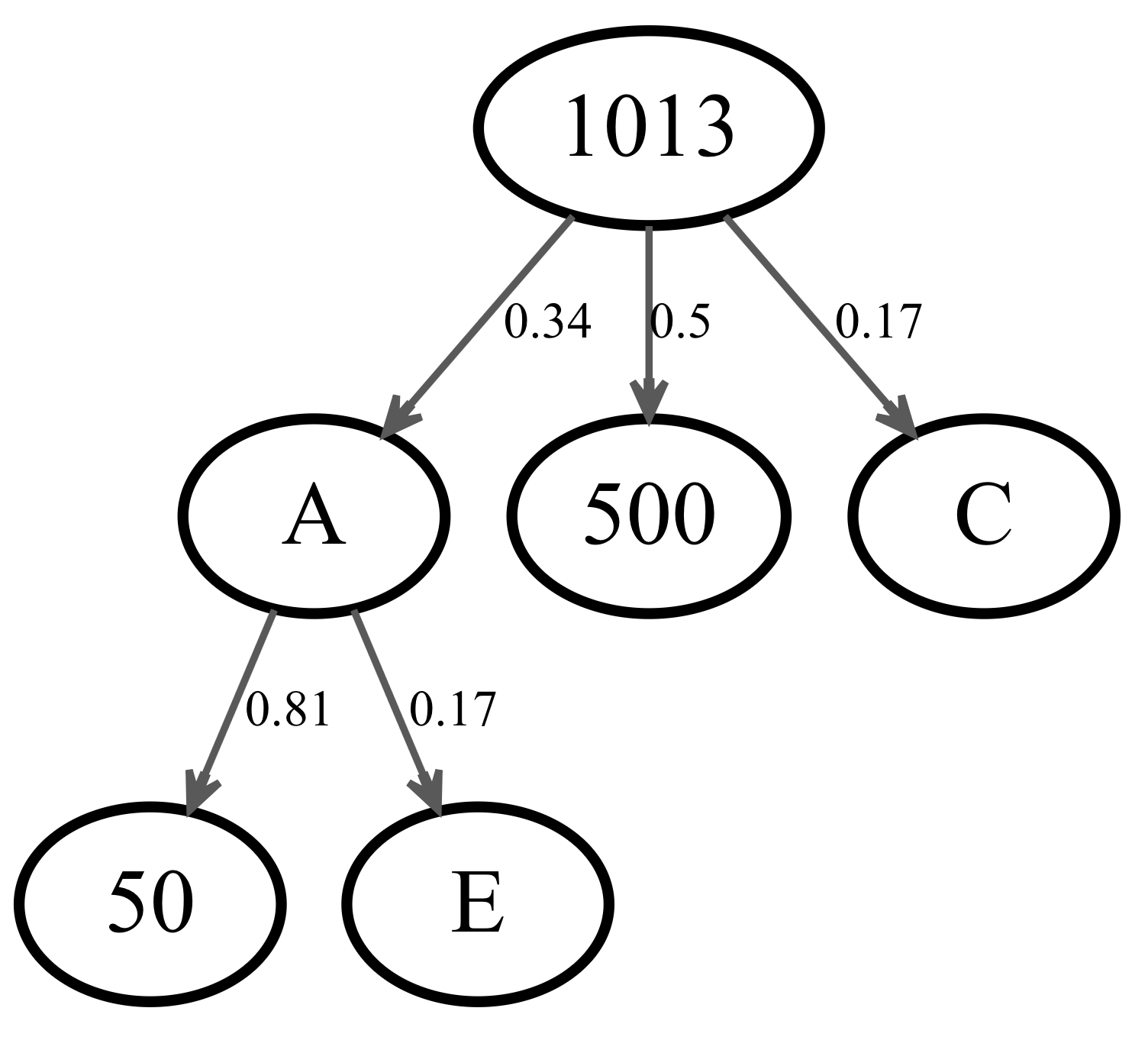}
	\caption{Example \texttt{countTree} output, using the tree defined in Figure \ref{fig:autotree}.  \texttt{countTree} is for use post-estimation to visualize tree data with root estimate provided by the WMM, and marginal counts shown within the corresponding known leaves.}
	\label{fig:countTreeex}
\end{figure}

The \texttt{wmmTree} function also modifies the \texttt{makeTree} object, so that path-specific root population size estimates and sampled probabilities are accessible via the \texttt{Get} function of the \textit{data.tree} package.  Two further tree rendering options post-estimation to aid visualization, \texttt{estTree} and \texttt{countTree}.  Both functions plot the tree structure with the final root estimate contained in the root and the average of branch samples along each sampled branch, but the former displays path-specific root estimates in the leaves while the latter displays known marginal counts in the leaves; tree renderings of the above example using \texttt{estTree} and \texttt{countTree} can be found in Figures \ref{fig:estTreeex} and \ref{fig:countTreeex}, respectively.  A summary of functions is given in Table \ref{table:autowmmfxns}, and further details can be found in the supplementary material.  

Example code to define the simple tree in Figure \ref{fig:autotree}, perform WMM estimation to determine the root population size estimate, and sample rendering code, is given in the following:
\begin{lstlisting}[]
# create admissible dataset
treeData <- data.frame("from" = c("Z", "Z", "Z", "A", "A"),
		"to" = c("A", "B", "C", "D", "E"),
		"Estimate" = c(4, 34, 1, 9, 1),
		"Total" = c(11, 70, 10, 10, 10),
		"Count" = c(NA, 500, NA, 50, NA),
		"Population" = c(FALSE, FALSE, FALSE, FALSE, FALSE),
		"Description" = c("First child of the root",
			"Second child of the root", "Third child of the root",
			"First grandchild", "Second grandchild"))
	
# make tree object using makeTree
tree <- makeTree(treeData)
	
# visualize tree pre-estimation
drawTree(tree)
	
# perform root node estimation
Zhats <- wmmTree(tree, sample_length = 15)
	
# print the estimates of the root node generated by the 15 iterations the weights 
# of each branch, and the final WMM estimate of the root node 
Zhats$estimates
Zhats$weights 
Zhats$root 
	
# prints the final rounded estimate of the root with confidence interval 
Zhats$uncertainty 
	
# show the average root estimate wiht 95% confidence interval, as well as average 
# estimates with confidence intervals for each node with a marginal count, 
# using data.tree function 'Get'
tree$Get('uncertainty')
	
# show the samples generated from each path which provides root estimates 
# using data.tree function 'Get'
tree$Get('targetEst_samples')
	
# show the probabilities sampled at each branch leading into the given node using 
# data.tree function 'Get'
tree$Get('probability_samples')
	
# draw tree post-estimation, with path-specfic estimates on respective  leaves,  
# root estimate in root node, and average branch probs on branches
estTree(tree)
	
# another tree drawn post-estimation, with marginal count in respective  leaves, 
# root estimate in root node, and average branch probs on branches
countTree(tree)
\end{lstlisting}


\subsection{Automated JAGS Modeling for Tree-Structure Data}\label{sec:autoJAGS}
The generation of JAGS code for tree-structured data has utility across many applications, and similar \text{R} packages have been developed to ease the implementation of other types of analysis, such as meta-analyses \cite{metaJAGS}.  Using MCMC with a Bayesian model in JAGS requires a bespoke modeling script dependent on the unique tree structure, limiting the accessibility of these methods due to the expertise required to successfully implement them.  To facilitate the use of Bayesian models for population estimation, this section outlines an easy-to-use automated method of generating JAGS models for tree-structured data in \texttt{R}. 

\subsubsection{\textit{JAGStree} Overview and Model Assumptions}\label{sec:autoJAGStheory}
As in Section \ref{sec:autowmmtheory}, we assume a basic structure to the data - information flows through a well-defined tree-like structure created by mutually exclusive data sources, and time is not a significant factor affecting reported data and estimation (see Section \ref{sec:discussion}.  We assume that expert or prior knowledge informs at least one marginal leaf count and some subset of tree branches, and sibling node counts follow $Multinomial$ prior distributions with the number of trials equal to the parent population and probabilities representative of branching proportions.  Prior branch probabilities themselves are assumed $Dirichlet$, and the root node is assigned either a \textit{LogNormal} or discrete \textit{Uniform} prior distribution.  Parameters of these distributions can be informed by survey data, as in Section \ref{sec:autowmmtheory}, but integration of data is not required to generate the code for a JAGS model, as it is instead specified when the model is implemented.   

\subsubsection{\textit{JAGStree} Implementation and Examples}\label{sec:autoJAGSimp}
The automated creation of a working JAGS model for tree-structured data as specified in Section \ref{sec:autoJAGStheory} has been written into the \texttt{R} package `\textit{JAGStree}' using \texttt{R} version 4.1.2 \cite{JAGStree}. 
The \textit{JAGStree} package functionality requires a specific data structure, which has been constructed to be compatible with the requirements of the \textit{AutoWMM} package.  However, the model-generating function, \texttt{makeJAGStree}, of the $JAGStree$ package only requires a data frame with columns named \textit{from} and \textit{to}.  These columns have the exact interpretation of those described in Section \ref{sec:autowmmtheory}, and are sufficient for defining the tree structure.

The \texttt{makeJAGStree} function generates a \texttt{.mod} or \texttt{.txt} file JAGS script for MCMC, which can then be independently implemented in \texttt{R} or \texttt{python}, for example.  The file begins with some preamble specifying the structural assumptions, followed by two chunks: a \texttt{data} chunk specifying the vectors of parameters required to inform branching distributions, and a \texttt{model} chunk describing the relationships among node and branching distributions.  As noted in Section \ref{sec:autoJAGStheory}, the root node is assumed to have either a \textit{LogNormal} prior distribution with parameters \texttt{mu} and \texttt{tau}, or discrete \textit{Uniform} distribution with bounds \texttt{Lz} and \texttt{Uz}.  JAGS does not have in-built discrete \textit{Uniform} functionality, so the model writes the prior to draw from a continuous \textit{Uniform} distribution, then rounds the sample to the nearest integer value.  \textit{LogNormal} samples are also rounded to the nearest integer value.  Once the root distribution is specified in the \texttt{model} chunk, the function then iteratively finds children of the current node, writing branching distribution parameters into the \texttt{data} chunk of the JAGS model, and relating branching distributions and node distributions within the \texttt{model} chunk.  For the sake of readability, if there are only two nodes in a sibling group, a $Beta/Binomial$ pairing of distributions is written instead of a $Dirichlet/Multinomial$, as the latter requires a marginal sampling approach to incorporate latent parameters (which are not supported by the built-in \texttt{multinomial} JAGS distribution). 

The naming scheme used is based on the tree's node labels in the \textit{from} and \textit{to} columns; for this reason, it is strongly recommended to name these nodes alphabetically or numerically, keeping longer descriptions to a \textit{Description} column in the data frame for cross-reference.  The root node name is extracted from the data, and branching probabilities are named according to the node in which the branch originates from, with a `\texttt{p}' prefix.  Sibling groups are treated as tuples, and named by condensing all sibling node names into one string; individual nodes of a sibling group are indexed by number, where the $n^{th}$ sibling refers to the $n^{th}$ node in the condensed name.  For example, for sibling group with nodes $A$, $B$, and $C$, the triple is referred to as \texttt{ABC} in the JAGS model, and node $B$ is given by \texttt{ABC[2]}.  The \texttt{data} chunk stores parameters used to inform branching probabilities, and these are named according to the branching probability they inform, with a `\texttt{.params}' suffix.  

To illustrate, we use the simple tree in Section \ref{sec:autowmmimp}, pictured in Figure \ref{fig:autotree}:
\begin{lstlisting}[]
# create admissible dataset
treeData <- data.frame("from" = c("Z", "Z", "Z", "A", "A"),
	"to" = c("A", "B", "C", "D", "E")))
	
# generate JAGS model in .mod or .txt file
makeJAGStree(data, prior = "lognormal", filename="sampleJAGS.mod")
\end{lstlisting}
The \texttt{makeJAGStree} function above generates the following JAGS model:
\begin{lstlisting}
# This JAGS model was created using 'makeJAGStree' in the 'JAGStree' package in R.
# The root may have lognormal or discretized uniform prior.
# Branching and leaf prior distributions are assumed Dirichlet and Multinomial, 
# respectively. 
	
data { 
	pZ.params <- c(pZ1, pZ2, pZ3); 
	pA.params <- c(pA1, pA2); 
} 
	
model { 
	Z.cont ~ dlnorm(mu, tau); 
	Z <- round(Z.cont); 
	pZ ~ ddirch(pZ.params); 
	Z.bin[1] <- Z; 
	pZ.bin[1] <- pZ[1]; 
	for (i in 2:3){ 
		Z.bin[i] <- Z.bin[i-1] - ABC[i-1] 
		pZ.bin[i] <- pZ[i]/(sum(pZ[i:3])) 
	} 
	for (i in 1:2){ 
		ABC[i] ~ dbinom(pZ.bin[i], Z.bin[i]) 
	} 
	ABC[3] <- Z.bin[1] - sum(ABC[1:2]); 
	
	pA ~ dbeta(pA.params[1], pA.params[2]); 
	DE[1] ~ dbinom(pA, ABC[1]); 
	DE[2] <- ABC[1] - DE[1]; 
	
} 
\end{lstlisting}
Once the \texttt{.mod} or \texttt{.txt} file has been generated, MCMC can proceed through a variety of methods, including via an \texttt{R} or \texttt{python} interface, such as $R2JAGS$ \cite{R2JAGS} or $PyJAGS$ \cite{PyJAGS}.  An example implementation of the \texttt{.mod} file using the $R2JAGS$ package in \texttt{R} is as follows:
\begin{lstlisting}[]
# load relevant packages
library(R2jags)
library(mcmcplots)
	
mcmc.data <- list( "DE" = c(50, NA),
                "ABC" = c(NA, 500, NA),
		# dirichlet parameters for prior on pZ
		"pZ1" = 4,     
		"pZ2" = 5,      
		"pZ3" = 1,    
		# beta parameters for prior on pA
		"pA1" = 10,      
		"pA2" = 1,       
		# lognormal mean Z
		"mu" = log(1000),    
		# lognormal precision (1/variance) of Z
		"tau" = 1/(0.1^2))    
	
# define parameters whose posteriors we are interested in
mod.params <- c("Z",  "ABC", "DE", "pZ", "pA")
	
# modify initial values
mod.inits.cont <- function(){list("Z.cont" = runif(1, 700, 1500),
		"ABC" = c(round(runif(1, 100, 200)), NA, NA),
		"pA" = as.vector(rbeta(1,1,1)),
		"pZ" = as.vector(rdirichlet(1, c(1,1,1))))
}
	
# Generate list of initial values to match number of chains
numchains <- 6
mod.initial.cont <- list()
i <- 1
while(i <= numchains){
	mod.initial.cont[[i]] <- mod.inits.cont()
	i = i+1
}
	
# now fit the model in JAGS and print output
mod.fit <- jags(data = mcmc.data, 
		inits = mod.initial.cont, 
		parameters.to.save = mod.params, 
		n.chains = numchains, 
		n.iter = 50000, n.burnin = 20000, 
		model.file = "sampleJAGS.mod")
	
print(mod.fit)
	
# optional plots using mcmcplots library
mod.fit.mcmc <- as.mcmc(mod.fit)
denplot(mod.fit.mcmc, parms = c("Z", "ABC[1]", "ABC[3]"))
denplot(mod.fit.mcmc, parms = c("pZ", "pA"))
traplot(mod.fit.mcmc, parms = c("Z", "ABC[1]", "ABC[3]"))
traplot(mod.fit.mcmc, parms = c("pZ", "pA"))
\end{lstlisting}
Additional simulation and plotting methods are available under many \texttt{R} and \texttt{python} libraries.  The \texttt{treeData} above can also be directly inserted into the \texttt{makeTree} function of the \textit{AutoWMM} package to render a descriptive tree; further \textit{AutoWMM} functionality, however, requires the other columns specified in Table \ref{table:autowmmdf}. 

\section{Discussion}\label{sec:discussion}
While the \textit{AutoWMM} package enables straightforward estimation of a root target population using the WMM, several extensions to the methodology could improve robustness of this method, and subsequently be included in the \texttt{R} package functionality.  For instance, while the WMM does have a number of Bayesian elements, one advantage of fully Bayesian models is the ability to account for error in marginal subpopulation counts.  This simple extension may improve root population size estimates where variance due to count uncertainty can be estimated.  Furthermore, the \textit{JAGStree} functionality overcomes one barrier of Bayesian model implementation on the tree topology, by writing executable JAGS modeling code \cite{flynnmethods}.  However, other essential steps to implementation remain, including choosing an appropriate root prior, assigning model parameters and data to appropriate components, and settings of the MCMC, such as initial conditions, number of chains, burn-in, etc.  Future work could expand the functionality to automate some of these tasks, though post-estimation analysis of model convergence and performance remains essential.  

The above models depend on the assumption that time is not a significant factor in reporting.  In some applications, a time-lag may exist in movement of data from root to leaf.  In this case, estimates of root node population size may not align with the time-period in which count data was collected at leaf nodes.  Additionally, when time-lag varies across informative paths, root population size estimates may appear to have artificially large variability.  Adjustments for filtration time could improve robustness of these methods in a wider variety of applications.

\section{Conclusion}\label{sec:conclusion}
The WMM and hierarchical Bayesian models each offer methods of population size estimation which synthesize evidence from available sources.  Model choice depends largely on balancing both the needs and constraints of the application at hand.  The \textit{AutoWMM} and \textit{JAGStree} \texttt{R} packages facilitate model creation and root population size estimation using both the WMM and hierarchical Bayesian models on trees, minimizing the barriers involved in implementing these methods.

\subsection*{Declaration of Interests}
The authors have no conflicts of interest to declare.

\subsection*{Acknowledgments}
This work was supported by NSERC Discovery Grant (RGPIN-2019-03957), an NSERC CGS-D, and a CIHR Doctoral Health System Impact Fellowship in partnership with the Public Health Agency of Canada (PHAC).

\bibliographystyle{WileyNJD-AMA}
\bibliography{statrefs}%


\end{document}